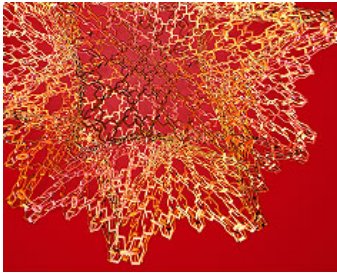

# Building Bricks with Bricks, with Mathematica

Pietro Codara[a], Ottavio M. D' Antona[a], Daniele Filaretti[b]

[a] Dipartimento di Informatica e Comunicazione, Università degli Studi di Milano
[b] Department of Computing, Imperial College London

## 1. Introduction

In this work we solve a special case of the problem of building an n-dimensional parallelepiped using a given set of n-dimensional parallelepipeds. Consider the identity

$x^3 = x(x-1)(x-2) + 3x(x-1) + x.$

For sufficiently large $x$, we associate with $x^3$ a cube with edges of size $x$, with $x(x-1)(x-2)$ a parallelepiped with edges $x, x-1, x-2$, with $3x(x-1)$ three parallelepipeds of edges $x, x-1, 1$, and with $x$ a parallelepiped of edges $x, 1, 1$. The problem we takle is the actual construction of the cube using the given parallelepipeds.

In [DDNP90] it was shown how to solve this specific problem and all similar instances in which a (monic) polynomial is expressed as a linear combination of a persistent basis. That is to say a sequence of polynomials $q_0 = 1$, and $q_k(x) = q_{k-1}(x)(x - r_k)$ for $k > 0$.

Here, after [Fil10], we deal with a multivariate version of the problem with respect to a basis of polynomials of the same degree (*binomial basis*). We show that it is possible to build the parallelepiped associated with a multivariate polynomial $P(x_1, \ldots, x_n) = (x_1 - s_1) \ldots (x_n - s_n)$ with integer roots, using the parallelepipeds described by the elements of the basis. We provide an algorithm in *Mathematica* to solve the problem for each $n$. Moreover, for $n = 2, 3, 4$ (in the latter case, only when a projection is possible) we use *Mathematica* to display a step by step construction of the parallelepiped $P(x_1, \ldots, x_n)$.

## 2. Theoretical Background

In this section we supply the mathematical description of the problem we deal with. Consider the relationship

$$p_n(x_1, \ldots, x_n) = \prod_{i=1}^{n}(x_i - s_i) = \sum_{S \subseteq \bar{n}} C_S \, q_S(x_1, \ldots, x_n) \qquad (1)$$

where $p_n$ is a monic *multilinear polynomial* with integer roots and the $2^n$ polynomials $q_S(x_1, \ldots, x_n)$, where $S \subseteq \bar{n} = \{1, 2, \ldots, n\}$, are a binomial basis (see below) of the vector space of multilinear polynomials.

- **The Binomial Basis**

Let, for $n > 0$, $\bar{n}$ denote the set $\{1, 2, ..., n\}$. With any subset $S$ of $\bar{n}$ we associate its *string function* $I_{\bar{n}}(S) = b_1 b_2 \cdots b_n$ where

$$b_i = \begin{cases} 0 & \text{if } i \notin S \\ 1 & \text{if } i \in S. \end{cases}$$

This allows us to refer to the subsets of $\bar{n}$ using a "bit string" notation. (The advantages of this approach will be clear in the following.) The definition of our basis exploits a complete binary tree $T_n = (V, E)$ of height $n > 0$ endowed with three labeling function: $q$, $b$ and $\phi$. The function

$$q : V \to \left\{ \frac{x_j + r}{j} : j \in N^+, r \in R \right\} \cup \{1\}$$

is defined as follows. Let $v$ be a node of $T_n$, $v_l$ its left child, and $v_r$ its right child. If $v$ is the root of $T_n$, then $q(v) = 1$, $q(v_r) = x_1 + 1$, and $q(v_l) = x_1$; otherwise, if $v$ has label $q(v) = \frac{x_j + r_j}{j}$ we define:

$$q(v_r) = \begin{cases} \frac{x_{j+1} + r_j + 1}{j+1} & \text{if } r_j > 0 \\ \frac{x_{j+1} + r_j + j}{j+1} & \text{otherwise}, \end{cases}$$

$$q(v_l) = \begin{cases} \frac{x_{j+1} + r_j - j}{j+1} & \text{if } r_j > 0 \\ \frac{x_{j+1} + r_j - 1}{j+1} & \text{otherwise}. \end{cases}$$

The second labeling function, $b : E \to \{0, 1\}$, is simply defined as follows: for each node $v \in V$ the edge connecting $v$ with its left child has label 1, while the edge connecting $v$ with its right child has label 0 (i.e. $b(v, v_l) = 1$ and $b(v, v_r) = 0$). The purpose of this function is to define a one-to-one correspondence between the maximal paths on $T_n$ and the subsets of $\bar{n} = \{1, ..., n\}$. Such correspondence allows us to use the following notation

$$q_S(x_1, ..., x_n), \; S \subseteq \bar{n}.$$

This way we can refer to the elements of the basis in terms of subsets instead of paths. Let now $p_n(x_1, ..., x_n) = (x_1 + s_1)(x_2 + s_2) \cdots (x_n + s_n)$ be an arbitrary multilinear polynomial. The third labeling function $\phi : V \to R$ is defined as follows. If $v \in V$ is the root of $T_n$, then $\phi(v) = 1$. Otherwise, if $q(v) = x_j + r$, we have

$$\phi(v) = \begin{cases} |r - j| + s_j & \text{if } r > 0 \\ |r + j| - s_j & \text{otherwise}. \end{cases}$$

We denote by $\alpha = v_1 \to v_2 \to \cdots \to v_k$, with $v_i \in V$ for each $i = 1, ..., k$, a path $\alpha$ in $T_n$. If $v_1$ is the root and $k = n$ (i.e. $v_k = v_n$ is a leaf), $\alpha$ is a maximal path in $T_n$. We can easily associate a binary string $b_\alpha$ of lenght $n$ with each maximal path $\alpha$ by defining the string:

$$b_\alpha = b(v_1, v_2) \ldots b(v_{n-1}, v_n).$$

We are now able to define the elements of our binomial basis as well as the coefficients of the linear combination expressing the polynomial $p_n$ in (1) in terms of such basis. In other words, given an arbitrary multilinear polynomial $p_n(x_1, ..., x_n)$ we are ready to explicitly write the right handside of (1).

For each $S \subseteq \bar{n}$, let $\alpha = v_1 \to v_2 \to \cdots \to v_n$ be the maximal path in $T_n$ such that $I_{\bar{n}}(S) = b_\alpha$, where. Then, we have:

$$q_S = q(v_1) \cdots q(v_n),$$
$$C_S = \phi(v_1) \cdots \phi(v_n).$$

The values $\phi(v)$ are not only a part of a product, but have an important meaning on their own. In fact, as we show in the next subsection, these values define the geometrical structure of the construction. Indeed, [Fil10] shows that this coefficients are a refinement of Eulerian numbers [Com74].

### ■ An Example

The following example shows the complete data structure for a generic polynomial $p_2(x_1, x_2) = (x_1 + s_1)(x_2 + s_2)$. In particular, the figure shows the complete binary tree $T_2$ with vertices and edges properly labeled.

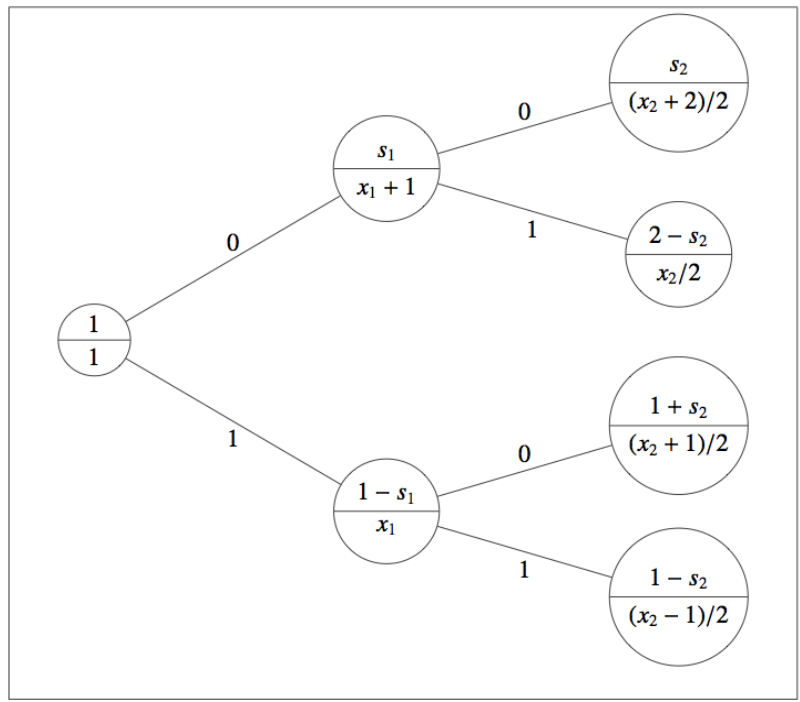

As previously mentioned we now consider each maximal path on $T_2$ in order to obtain both the elements of the basis and the coefficient that links $p_2$ with this basis. The following table shows the results of the computation.

| S | $I_{\bar{n}}(S)$ | $q_S$ | $C_S$ |
|---|---|---|---|
| ∅ | 00 | $(x_1 + 1) \cdot (x_2 + 2)/2$ | $s_1 s_2$ |
| {2} | 01 | $(x_1 + 1) \cdot x_2/2$ | $2s_1 - s_1 s_2$ |
| {1} | 10 | $x_1 \cdot (x_2 + 1)/2$ | $1 - s_1 + s_2 - s_1 s_2$ |
| {1, 2} | 11 | $x_1 \cdot (x_2 - 1)/2$ | $1 - s_1 - s_2 + s_1 s_2$ |

## Construction Algorithm

The algorithm we are going to show takes an arbitrary multilinear polynomial $p_n(x_1, \ldots, x_n) = (x_1 + s_1)(x_2 + s_2) \cdots (x_n + s_n)$ as an input and builds the associated $n$-dimensional parallelepiped using the smaller $n$-dimensional parallelepipeds associated with the elements of the binomial basis of degree $n$.

An auxiliary function sets up the enviroment (i.e. it sets up some variables we will need in the actual construction) and then calls the recursive procedure Draw, that actually performs the "hard work".

```
input :
The complete binary tree T_n endowed with the labeling functions q and ϕ;
a set of values x_1, ..., x_n.

output : construction of the n dimensional
```

```
parallelepiped associated with p_n with the parallelepipeds
associated with the elements of the binomial basis of degree n.

origin ← [0,...,0] (* n-times *)
edges ← [0,...,0] (* n-times *)
v ← the root of T_n
Draw (v)
```

The function Draw recursively visits the tree $T_n$, meanwhile producing the required construction. It works on the assumption that our data structure (i.e. the tree $T_n$ endowed with our three labeling functions) not only allows us to define the element of the binomial basis and relative coefficients, but also contains complete information about the geometrical structure of the object we are going to build.

The basic idea of the algorithm is as follow. Let $v$ be a vertex of $T_n$, and let $h(v)$ be its height. Then: (i) if $h(v) = n$ (i.e. $v$ is a leaf) then $v$ represents the parallelepiped associated with the maximal path ending in $v$; (ii) if $h(v) < n$ then $v$ represents the part of construction we obtain by aligning on the $(n - h(v))$-th coordinate the part of construction associated with its left child (repeated $\phi(v_l)$ times) with the part of construction associated with its right child (repeated $\phi(v_r)$ times).

The procedure works by exaustively visiting $T_n$ and using the informations provided by each node in order to build the object, following the principles explained in the previous paragraph.

```
if h(v) = 0 then
    for i ← 1 to |ϕ(v_l)| do
        draw(v_l);
    for i ← 1 to |ϕ(v_r)| do
        draw(v_r);

if h(v) = i then
    edges[n - i + 1] ← q(v)(x_i);
    for i ← 1 to |ϕ(v_l)| do
        draw(v_l);
    for i ← 1 to |ϕ(v_r)| do
        draw(v_r);
    for j ← i + 1 to n do
        origin[j] ← 0;

if h(v) = n then
    edges[1] ← q(v)(x_n);
    drawBrick(origin, edges);
    origin[1] ← origin[1] + edges[1];
```

# 3. Functions Tour

In this section we describe the *Mathematica* implementation of the main algorithm (the one that actually provides a step by step construction of the parallelepiped) and of the other functions and procedures. Our implementations is built upon a set of (small) *Mathematica* functions, which we can easily classify in three main categories.

▪ **Auxiliary Functions**

The following utility functions are frequently used in other parts of the implementation.

| Syntax | getAllSubset[*n*] |
|--------|-------------------|

| **Input** | an integer *n* |
| --- | --- |
| **Output** | a list containing all the subsets of {1, ..., *n*} |

```
getAllSubsets[n_] := (
   set = Table[i, {i, n}];
   subsets = Subsets[set];
   Return[subsets];
   );
```

| **Syntax** | **fromSubsetToBinary**[*S*, *n*] |
| --- | --- |
| **Input** | an integer *n*; a subset $S \subseteq \bar{n}$ |
| **Output** | binary string representation of the given subset |

```
fromSubsetToBinary[sub_, n_] := (
   bitString = Table[0, {i, n}];
   For[i = 1, i ≤ n, i++,
    If[Count[sub, i] == 0,
       bitString[[i]] = 0,
       bitString[[i]] = 1
      ];
    ];
   Return[bitString];
   );
```

| **Syntax** | **getAllBinarySubsets**[*m*] |
| --- | --- |
| **Input** | an integer *m* |
| **Output** | a list containing the binary string representation of all the subsets of {1, ..., *m*}. Note that this result is actually a list containing all binary strings of lenght *m* |

```
getAllBinarySubsets[m_] := (
   subs = getAllSubsets[m];
   l = Length[subs];
   binSubs = Table[0, {j, l}];
   For[j = 1, j ≤ l, j++,
    binSubs[[j]] = fromSubsetToBinary[subs[[j]], m];
    ];
   Return[Sort[binSubs]];
   );
```

| **Syntax** | **mySign**[*n*] |
| --- | --- |
| **Input** | an integer *n* |
| **Output** | The sign of *n*, with the exception that zero is treated as negative |

```
mySign[n_] := (
   If [n ≠ 0, Return[Sign[n]]];
   If [n == 0, Return[-1]];
   );
```

### ■ Tree Manipulation Functions

The following functions allows to solve the problem both algebrically and geometrically, exploiting the complete binary tree structure.

| **Syntax** | generateTree[n, s] |
| --- | --- |
| **Input** | an integer *n* > 1 and a vector *s* of lenght *n* containing the roots of the polynomial |

p($x_1$, ..., $x_n$) = ($x_1$ + s[1]) ... ($x_n$ + s[n])
**Output**     the data structure defining $T_n$
**Notes**     The complete binary tree is implemented using an array of nodes. We point out that this structure is meant to be manipulatd by the other procedures, not by the user.
    Here is the structure of each node:
        [1]: left child index
        [2]: right child index
        [3]: $\phi$
        [4]: root of the associated polynomial (see definition of labeling function $q$)
        [5]: height of the node

```
generateTree[n_, s_] := (
   (* Init the data structure *)
   node = {};
   (* Set up first node (root) *)
   root = {-1, -1, 1, 0, 0};
   node = Append[node, root];
   (* Now we start to build the tree *)
   i = 0;
   While[Length[node] < 2^(n+1) - 1,
    (* for each node, we compute the attributes of its (2) childs *)
    j = node[[i+1]][[5]] + 1; (* index x_j*)
    If[node[[i+1]][[4]] >= 0,
     rDx = node[[i+1]][[4]] + 1; (* r: radice *)
     mulDx = Abs[node[[i+1]][[4]] - node[[i+1]][[5]]] + s[[j]];
     rSx = node[[i+1]][[4]] - (node[[i+1]][[5]]);
     mulSx = Abs[node[[i+1]][[4]] + 1] - s[[j]],
     (* else *)
     rDx = node[[i+1]][[4]] + (node[[i+1]][[5]]);
     mulDx = Abs[node[[i+1]][[4]] - 1] + s[[j]];
     rSx = node[[i+1]][[4]] - 1;
     mulSx = Abs[node[[i+1]][[4]] + node[[i+1]][[5]]] - s[[j]];
    ];
    (* here we actually "create" the node *)
    figlioSx = {-1, -1, mulSx, rSx, j};
    figlioDx = {-1, -1, mulDx, rDx, j};
    (* ... and put it in the list*)
    node = Append[node, figlioSx];
    node = Append[node, figlioDx];
    (* finally,
    we update the current node (it must contains pointers to its childs) *)
    node[[i+1]][[1]] = Length[node] - 1;
    node[[i+1]][[2]] = Length[node] ;
    i = i + 1;
   ];
   Return[node];
  );
```

| | |
|---|---|
| **Syntax** | getRootsAndCoefs[$T_n$] |
| **Input** | a data structure representing $T_n$ (usually generated by the function generateTree[]) |
| **Output** | a list (of lists) containing the roots of the $2^n$ elements of the basis; a list containing the coefficients between the polynomial $p$ and the elements of the basis |

```
getRootsAndCoefs[tree_] := (
   n = Log[2, Length[tree] + 1] - 1;
   mulVector = Table[0, {i, n}];
   rootVector = Table[0, {i, n}];
   roots = {};
   coef = {};
   visitAndExtract[tree, 2];
   visitAndExtract[tree, 3];
   Return[{roots, coef}];
   );
```

The following procedure is meant to be called exclusively by the other procedure above (getRootsAndCoefs).

```
visitAndExtract[tree_, node_] := (
   nodeType = tree[[node]][[5]];
   mulVector[[nodeType]] = tree[[node]][[3]];
   rootVector[[nodeType 1]] = tree[[node]][[4]];
   If [nodeType == n,
     (* current node is a lead *)
     roots = Append[roots, rootVector];
     coef = Append[coef, Product[mulVector[[i]], {i, 1, Length[mulVector]}]],
     (* current node is not a leaf *)
     visitAndExtract[tree, tree[[node]][[1]]];
     visitAndExtract[tree, tree[[node]][[2]]];
     ];
   );
```

| | |
|---|---|
| **Syntax** | drawBrick[$T_n$, node, indeterminates] |
| **Input** | The tree $T_n$, as generated by the generateTree procedure; the starting node (usually the root) index. |
| **Output** | The final construction, in the form of a list of Cuboid objects. |

```
(* This procedure is only for convenience. The
 "hard work" is undertaken by the "visit" procedure *)
drawBrick[tree_, node_, soso_] := (
   (*colors = {Red, Green, Blue, Black, White,
       Gray, Cyan, Magenta, Yellow, Brown, Orange, Pink, Purple};*)
   colors = {White, Blue, Yellow, Pink, Green, Red, Cyan, Magenta};
   zyx = soso;
   lx = ly = lz = 0;
   x0 = y0 = z0 = 0;
   fig = {};
   n = Log[2, Length[tree] + 1] - 1;
   Return[visit[tree, node]];
   );
```

The following procedure implements the algorithm that actually "builds bricks with bricks". Of course, this will only work for $n = 1, 2, 3$, and, in some special cases, $n = 4$. However, the full algorithm [Fil10] is valid for each $n$.

```
visit[tree_, node_] := (
   (* first, we find out the height of the node *)
   nodeType = tree[[node]][[5]];

   (* now, we have different possibilities... *)
   Switch[nodeType,
    (* t *)
    n - 3,
    (* visit childs *)
    For[i = 1, i ≤ tree[[tree[[node]][[1]]]][[3]], i++,
     visit[tree, tree[[node]][[1]]];];
    For[i = 1, i ≤ tree[[tree[[node]][[2]]]][[3]], i++,
     visit[tree, tree[[node]][[2]]];],

    (* z *)
    n - 2,
    lz = zyx[[1]] + tree[[node]][[4]] ;
    (* visit childs *)
    For[i = 1, i ≤ tree[[tree[[node]][[1]]]][[3]], i++,
     visit[tree, tree[[node]][[1]]];];
    For[i = 1, i ≤ tree[[tree[[node]][[2]]]][[3]], i++,
     visit[tree, tree[[node]][[2]]];];
    x0 = 0;
    y0 = 0;
    z0 = z0 + lz,

    (* y *)
    n - 1,
    (* compute lenght of x edge *)
    ly = (zyx[[2]] + tree[[node]][[4]]) / 2;
    (* visit childs *)
    For[l = 1, l ≤ tree[[tree[[node]][[1]]]][[3]], l++,
     visit[tree, tree[[node]][[1]]];];
    For[l = 1, l ≤ tree[[tree[[node]][[2]]]][[3]], l++,
     visit[tree, tree[[node]][[2]]];];
    x0 = 0;
    y0 = y0 + ly,

    (* x *)
    n,
    lx = (zyx[[3]] + tree[[node]][[4]]) / 3 ;
    fig = Append[fig, RandomChoice[colors]];
    fig = Append[fig, Cuboid[{x0, y0, z0}, {x0 + lx, y0 + ly, z0 + lz}]];
    (* update position *)
    x0 = x0 + lx;
   ];
   Return[fig];
  );
```

## 4. Examples

In this sections we provide complete examples of execution of our main algorithm. We will consider two examples: a 3D one, and a special 4D example in which we will use a projection to output a 3D draw.

### ▪ A Three-dimensional Example

```
Clear[n, tree, temp, roots, coefs, radici, p];
```

As usually, we fix the roots $s_1$, $s_2$ and $s_3$ of the polynomial/parallelepiped $p$ that we want to build using the elements of the binomial basis. In this example we have $p(x_1 \, x_2 \, x_3) = x_1 \, x_2 \, x_3$ and hence $s_1 = s_2 = s_3 = 0$.

```
s = {0, 0, 0};
```

We then fix arbitrary values for $x_1$, $x_2$ and $x_3$ (but remember to pay attention to negative edges).

```
lzyx = {3, 3, 3};
```

Finally we create the complete binary tree $T_3$ with all the required labeling functions, and print a tabular representation of it.

```
T3 = generateTree[3, s];
TableForm[T3, TableHeadings → {None, {"vs", "vd", "ϕp", "r", "h"}},
  TableSpacing → {1, 3}, TableAlignments → Right]
```

| $v_s$ | $v_d$ | $\phi_p$ | r | h |
|---:|---:|---:|---:|---:|
| 2 | 3 | 1 | 0 | 0 |
| 4 | 5 | 1 | 0 | 1 |
| 6 | 7 | 0 | 1 | 1 |
| 8 | 9 | 1 | -1 | 2 |
| 10 | 11 | 1 | 1 | 2 |
| 12 | 13 | 2 | 0 | 2 |
| 14 | 15 | 0 | 2 | 2 |
| -1 | -1 | 1 | -2 | 3 |
| -1 | -1 | 2 | 1 | 3 |
| -1 | -1 | 2 | -1 | 3 |
| -1 | -1 | 1 | 2 | 3 |
| -1 | -1 | 1 | -2 | 3 |
| -1 | -1 | 2 | 1 | 3 |
| -1 | -1 | 3 | 0 | 3 |
| -1 | -1 | 0 | 3 | 3 |

We are now ready to ask our algorithm to implement the actual construction. As previously mentioned, the actual construction is made via a list of *Mathematica's* Cuboid objects.

```
costruzione = drawBrick[T3, 1, lzyx];
TableForm[costruzione]
```

```
RGBColor[0, 1, 1]
Cuboid[{0, 0, 0}, {1/3, 1, 3}]
RGBColor[0, 0, 1]
Cuboid[{1/3, 0, 0}, {5/3, 1, 3}]
RGBColor[0, 1, 0]
Cuboid[{5/3, 0, 0}, {3, 1, 3}]
RGBColor[0, 0, 1]
Cuboid[{0, 1, 0}, {2/3, 3, 3}]
RGBColor[1, 0.5, 0.5]
Cuboid[{2/3, 1, 0}, {4/3, 3, 3}]
RGBColor[1, 0, 0]
Cuboid[{4/3, 1, 0}, {3, 3, 3}]
```

As a final step, we can simply ask *Mathematica* to plot our list of parallelepipeds.

```
Graphics3D[costruzione, Axes → False, Boxed → False]
```

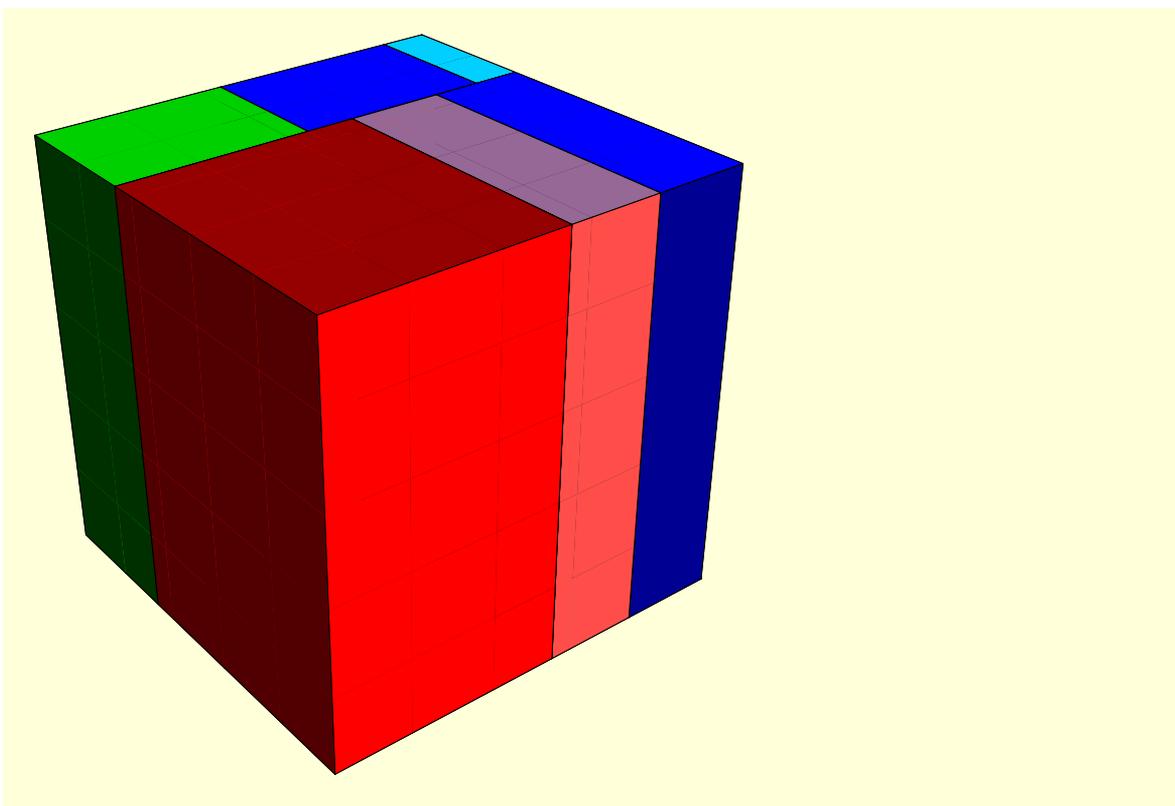

### ◾ A Four-dimensional Example

In this example we show how it is possible to provide a graphical construction for a (particular) four-dimensional case. More precisely, we are going to construct (as usually, in term of the elements of the binomial basis) the polynomial $p = x_1 \, x_2 \, x_3 \, x_4$ exploiting the fact that all the parallelepipeds associated with the elements of the basis have the fourth edge of the same lenght: this allows us to implement a projection.

```
Clear[n, tree, temp, roots, coefs, radici, p];
```

As usually, we choose values for the roots of $p$

```
s = {0, 0, 0, 0};
```

and values for $x_1$, $x_2$, $x_3$ and $x_4$.

```
lzyx = {4, 4, 4, 4};
```

We are now ready to generate the data structure $T_4$.

```
T4 = generateTree[4, s] ;
TableForm[T4, TableHeadings → {None, {"vs", "vd", "φp", "r", "h"}},
  TableSpacing → {1, 3}, TableAlignments → Right]
```

| $v_s$ | $v_d$ | $\phi_p$ | r | h |
|---:|---:|---:|---:|---:|
| 2 | 3 | 1 | 0 | 0 |
| 4 | 5 | 1 | 0 | 1 |
| 6 | 7 | 0 | 1 | 1 |
| 8 | 9 | 1 | -1 | 2 |
| 10 | 11 | 1 | 1 | 2 |
| 12 | 13 | 2 | 0 | 2 |
| 14 | 15 | 0 | 2 | 2 |
| 16 | 17 | 1 | -2 | 3 |
| 18 | 19 | 2 | 1 | 3 |
| 20 | 21 | 2 | -1 | 3 |
| 22 | 23 | 1 | 2 | 3 |
| 24 | 25 | 1 | -2 | 3 |
| 26 | 27 | 2 | 1 | 3 |
| 28 | 29 | 3 | 0 | 3 |
| 30 | 31 | 0 | 3 | 3 |
| -1 | -1 | 1 | -3 | 4 |
| -1 | -1 | 3 | 1 | 4 |
| -1 | -1 | 2 | -2 | 4 |
| -1 | -1 | 2 | 2 | 4 |
| -1 | -1 | 2 | -2 | 4 |
| -1 | -1 | 2 | 2 | 4 |
| -1 | -1 | 3 | -1 | 4 |
| -1 | -1 | 1 | 3 | 4 |
| -1 | -1 | 1 | -3 | 4 |
| -1 | -1 | 3 | 1 | 4 |
| -1 | -1 | 2 | -2 | 4 |
| -1 | -1 | 2 | 2 | 4 |
| -1 | -1 | 1 | -3 | 4 |
| -1 | -1 | 3 | 1 | 4 |
| -1 | -1 | 4 | 0 | 4 |
| -1 | -1 | 0 | 4 | 4 |

Now we ask the function drawBrick to provide us the with the actual construction. Notice that, in order to allow this example to work, we used a little trick: the function drawBrick does not start executing (as usually) on the root of $T_4$, but instead it start executing on its left subtree. This is possible only in a few special cases where one subtree is "empty" (i.e. all the values for $\phi$ are zero).

```
costruzione = drawBrick[T4, 2, lzyx];
TableForm[costruzione]
```

```
RGBColor[1, 0, 1]
Cuboid[{0, 0, 0}, {1/3, 1, 3}]
RGBColor[1, 1, 0]
Cuboid[{1/3, 0, 0}, {2, 1, 3}]
RGBColor[1, 0, 0]
Cuboid[{2, 0, 0}, {11/3, 1, 3}]
RGBColor[0, 1, 0]
Cuboid[{11/3, 0, 0}, {16/3, 1, 3}]
RGBColor[1, 0.5, 0.5]
Cuboid[{0, 1, 0}, {2/3, 7/2, 3}]
RGBColor[0, 1, 1]
Cuboid[{2/3, 1, 0}, {4/3, 7/2, 3}]
RGBColor[0, 1, 0]
Cuboid[{4/3, 1, 0}, {10/3, 7/2, 3}]
RGBColor[1, 0, 1]
Cuboid[{10/3, 1, 0}, {16/3, 7/2, 3}]
RGBColor[1, 0, 0]
Cuboid[{0, 7/2, 0}, {2/3, 6, 3}]
RGBColor[1, 0, 0]
Cuboid[{2/3, 7/2, 0}, {4/3, 6, 3}]
GrayLevel[1]
Cuboid[{4/3, 7/2, 0}, {10/3, 6, 3}]
GrayLevel[1]
Cuboid[{10/3, 7/2, 0}, {16/3, 6, 3}]
RGBColor[1, 1, 0]
Cuboid[{0, 0, 3}, {2/3, 3/2, 8}]
RGBColor[0, 0, 1]
Cuboid[{2/3, 0, 3}, {4/3, 3/2, 8}]
RGBColor[1, 0, 0]
Cuboid[{4/3, 0, 3}, {10/3, 3/2, 8}]
GrayLevel[1]
Cuboid[{10/3, 0, 3}, {16/3, 3/2, 8}]
RGBColor[1, 0, 0]
Cuboid[{0, 3/2, 3}, {2/3, 3, 8}]
RGBColor[0, 1, 1]
Cuboid[{2/3, 3/2, 3}, {4/3, 3, 8}]
RGBColor[1, 0.5, 0.5]
Cuboid[{4/3, 3/2, 3}, {10/3, 3, 8}]
GrayLevel[1]
Cuboid[{10/3, 3/2, 3}, {16/3, 3, 8}]
RGBColor[1, 0.5, 0.5]
Cuboid[{0, 3, 3}, {1, 6, 8}]
RGBColor[0, 1, 0]
Cuboid[{1, 3, 3}, {2, 6, 8}]
RGBColor[1, 0.5, 0.5]
Cuboid[{2, 3, 3}, {3, 6, 8}]
RGBColor[0, 1, 0]
Cuboid[{3, 3, 3}, {16/3, 6, 8}]
```

Finally, we display the result.

```
Graphics3D[costruzione, Axes → False, Boxed → False]
```

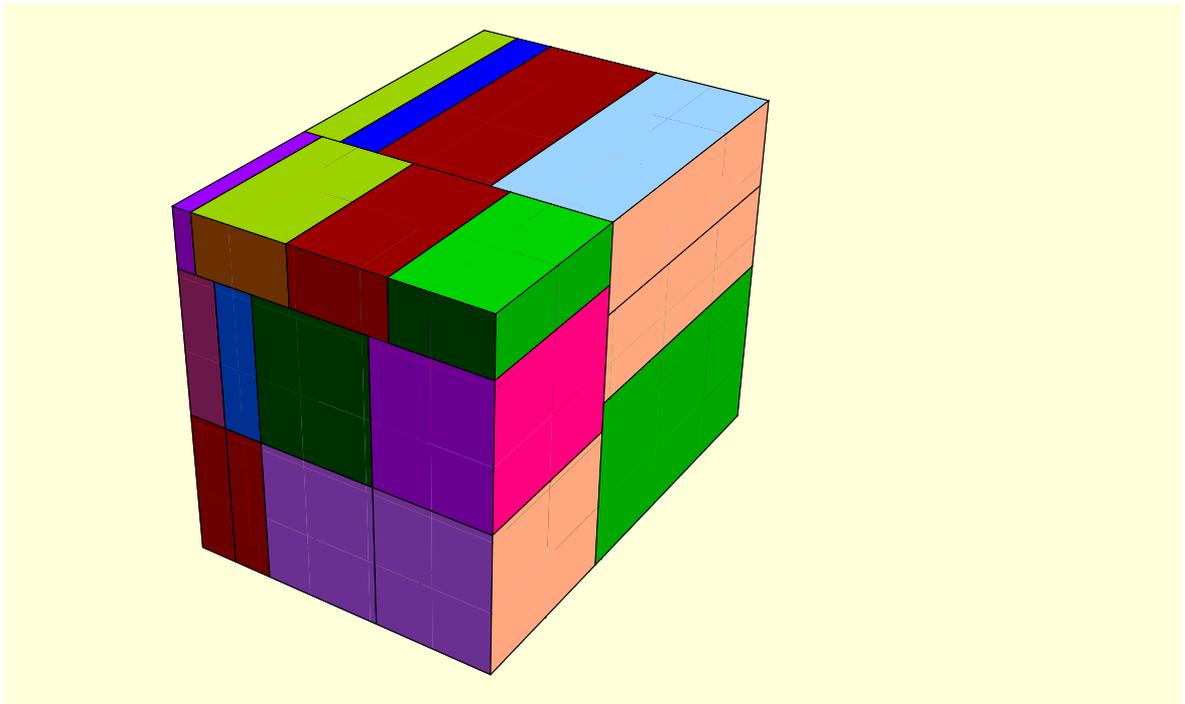